\documentstyle[preprint,prb,aps]{revtex}
\tightenlines
\def\nb{{{\bar n}}}
\def\rr{{{\bf r}}}
\def\rrp{{{\bf r^\prime}}}
\def\nup{{{n_\uparrow}}}
\def\ndo{{{n_\downarrow}}}

\begin{document}
\draft

\title{Stabilized Spin-Polarized Jellium Model \\ and\\ Odd-Even
Alternations in Jellium Metal Clusters}

\author{M.~Payami and N.~Nafari}
\address{Center for Theoretical Physics and Mathematics,
 P.~O.~Box 11365-8486, Tehran, Iran;\\ and\\
International Center for Theoretical Physics, P.~O.~Box 586, 34100 Trieste,
Italy}

\date{\today}
\maketitle
\begin{abstract}
In this paper, we have considered the mechanical stability of a jellium
system in the presence of spin degrees of freedom and have generalized
the stabilized jellium model, introduced by J.~P.~Perdew, H.~Q.~Tran,   and
E.~D.~Smith [~Phys.~Rev.~B{\bf~42},~11627~(1990)], to a spin-polarized case.
By applying this generalization to metal clusters (Al, Ga, Li, Na, K, Cs), we
gain
additional insights about the odd-even alternations, seen in their ionization
potentials. In this
generalization, in addition to the electronic degrees of freedom, we allow the
positive jellium background to expand as the clusters' polarization increases.
In fact, our
self-consistent calculations of the energetics of alkali metal clusters with
spherical geometries, in the context of density functional
theory and local spin density approximation, show that the energy of a
cluster
is minimized for a configuration with maximum spin compensation (MSC). That is,
for
clusters with even number of electrons, the energy minimization gives rise to
complete compensation ($N_\uparrow=N_\downarrow$), and for clusters with odd
number of electrons, only one electron remains uncompensated
($N_\uparrow-N_\downarrow=1$). It is this MSC-rule which
 gives rise to alternations in the ionization potentials. Aside from very few
exceptions, the MSC-rule is also at work for other metal culsters (Al, Ga) of
various sizes.
\end{abstract} \pacs {36.40, 71.10, 31.15.E}

\section{Introduction}
\label{sec1}

The subject of metal clusters has gained considerable momentum in recent years
\cite{ekardt,knight,deheer87,deheer93,brack93,yano93}.
The first step for studying these systems, has been to employ
the spherical jellium model (JM) along with the density functional formalism
(DFF)\cite{hohen64,kohnsham65,parr89,dreizler90}.
However, the
spherical-JM despite its initial successes suffers from two main deficiencies.

Firstly, the approximation that metal clusters assume spherical geometry, can
only be justified for large closed-shell clusters, but not
for very small clusters having closed- or open-shell electronic
configurations. Therefore, some authors have used the deformed spheroidal- or
ellipsoidal-JMs
\cite{clem,ekpenz88,ekpenz91,penzek90,laur,hirsch94,kohl,yano95}.
The results of {\it ab initio} molecular dynamics
calculations \cite{bonacic} confirm the overall shapes predicted by the
deformed-JMs.
Koskinen and coworkers
\cite{koskinen} in their ultimate-JM have assumed the jellium background to
be completely deformable both in shape and in density. However, the ultimate-JM
is applicable only when the local density parameter, $r_s$, does not differ
much from 4.18, i.e., the $r_s$-value for which the jellium system is in
mechanical equilibrium. In the following we will discuss the mechanical
stability of the JM.

The second drawback arise from the JM itself. It is well-known that the
JM yields negative surface energies \cite{lang70} at high electron densities
$(r_s\le
2)$, and negative bulk moduli \cite{ashlang67} for $r_s\approx 6$. These
drawbacks are expected to manifest themselves in the jellium metal clusters
too. To overcome these deficiencies,
 some authors have brought in the ionic structure either
perturbatively
\cite{lang70,ashlang67,lang71,manninen86,schone} or
variationally \cite{monp,monplan,sahni}.
However, other researchers have tried to modify the JM
in such a way as to keep its simplicity and yet avoid the above-mentioned
drawbacks\cite{diazshore84,shore91,rose91,diazshore89,pertran}.
These authors emphasize that the jellium system is not in mechanical
equilibrium, except for $r_s\approx 4.18$.
In particular, we refer to the work of Perdew, Tran, and Smith\cite{pertran}
who have
introduced the stabilized jellium model (SJM).
Applications of the SJM to infinite and semi-infinite simple metals
\cite{pertran,fiol95} yield realistic estimates for the cohesive and surface
energies.
In fact, the SJM-calculations of the energetics of simple metal
clusters with spherical geometries \cite{braj93}, and voids
\cite{ziesche}
 show an appreciable improvements over the
simple JM results. For example, the energy per particle for large sodium metal
clusters changes from $\sim -2 eV$ to $\sim -6eV$.

However, according to what was mentioned earlier, the spherical-SJM may only
be suitable for clusters with closed-shell electronic configurations. The
reasons
are two-fold: i) In the SJM, the stabilization has been accomplished for a
spin-compensated system which is not necessarily suitable for open-shell
clusters. ii) The electronic charge densities for open-shell systems do not
have spherical symmetry.

Thus, to study open-shell clusters, two modifications over the spherical-SJM
need be done. Firstly, open-shell clusters are not generally expected to have
zero polarization. This is particularly true for clusters with odd number of
valence electrons. Therefore, the SJM must be generalized to the stabilized
spin-polarized jellium model (SSPJM) which is the subject of this work.
Secondly, because of the deformations due to the Jahn-Teller effect\cite{jahn}
for open-shell systems, one should employ deformed nonspherical shapes
for the jellium background.
R\"othlisberger and Andreoni
\cite{roth}
by using the Car-Parrinello \cite{car} method, i.e., the unified
DFF and
molecular dynamics, have obtained interesting results. The results of their
extensive computer simulation show that on the one hand, the overall shapes of
clusters
change when the number of atoms in clusters change. This supports the
deformed jellium models. On the other hand, they show that the average distance
between the
nearest neighbors, $d_{av}$, in sodium microclusters alternate with increasing
number of sodium atoms (see Fig.~15 of Ref.~[\ref{roth}]).
We will show that these alternations are predicted via the SSPJM.

Guided by these results, we have asked whether it is possible,
to keep the spherical geometry and further improve the
results obtained using the spherical-SJM.
 Our answer is positive \cite{payam}. In fact, in the spherical-SSPJM, by
allowing the volume of a cluster change as a function of its
spin-polarization, we arrive at new results which are absent in the previous
spherical-JM or spherical-SJM and gain further insights for the observed
odd-even
alternations in the ionization potentials (IP) of metal clusters
\cite{deheer93,kappes,homer}. For clarification purposes, we mention that it is
the
difference of the level separations near the Fermi level for the neutral and
ionized clusters which gives rise to alternations in IPs. These changes can
show up: i) By allowing nonspherical shape deformations of the jellium
background (JB) while keeping its volume fixed
\cite{clem,ekpenz88,ekpenz91,penzek90,laur,hirsch94,kohl,yano95}.
 ii) By preserving the spherical shape of the JB and allowing its
volume per valence electron change. The latter is what we are addressing in
this paper. We
will see that the mechanism causing changes in the JB volume is rooted in the
stabilization of the spin-polarized JM. At any rate, in reality,
both effects of nonspherical deformations and volume changes of the JB are in
operation.

Our self-consistent SSPJM-calculations of the
energetics of metal clusters (Al, Ga, Li, Na, K, Cs) show that the total energy
of a cluster
is minimized for a configuration with maximum spin compensation (MSC). That is,
for clusters with even number of electrons, the energy minimization
gives rise to complete compensation ($N_\uparrow=N_\downarrow$),
and for
clusters with odd number of electrons, only one electron remains uncompensated
($N_\uparrow-N_\downarrow=1$). According to our calculations, the only
exceptions to the MSC-rule for both neutral and ionized metal clusters (Al,
Ga, Li, Na, K, Cs) of various sizes ($2\le N\le 42$) are Al$_{12}$,
Al$^+_{13}$, Al$_{14}$, and Ga$_{12}$. Here, $N$ is the total number of
valence electrons and $N/z=n$ is the number of $z$-valent atoms in a cluster.
The MSC-rule together with
the monotonically  increasing variation of the cluster size as a function of
spin polarization  give rise to the alternation of the mean
distances between the nearest neighbors as a function of cluster size, $N$.
As a result of these alternations, the total
energies change, and thereby, the IPs alternate.
In this work, by taking a diffusion layer for the jellium \cite{alon}, we have
 also repeated our calculations for Na-clusters.
 Application of the diffuse-stabilized spin-polarized jellium model
(dif-SSPJM) has
lowered our calculated ionization energies, and as a result, we have obtained
closer agreement with experimental data.

It would be interesting to compare the results of our calculations, i.e., the
MSC-rule, with the results of the previous calculations in which the density
parameter, $r_s$, was assumed to be fixed for all cluster sizes. The situation
in those models is similar to that of atoms in which the external
potential produced by the positive charge background with given $r_s$-value is
fixed for different electronic configurations and only the electrons are
allowed to redistribute themselves. We note
that in the case of atoms, clearly the external potential due to nuclear charge
remains fixed. In these cases according to the Pauli exclusion principle,
electrons with parallel spins are kept apart, i.e., further apart as compared
with the electrons having antiparallel spins. In fact, this is the way the
electrons reduce their total electrostatic energies and this is why the
electrons in a given shell assume maximum polarization consistent with the
Pauli exclusion principle. In other words, the Hund's first rule is applicable
in these cases. On the other hand, in the case of the SSPJM, the total energy
of the system is reduced by allowing the cluster radii to expand and the ions
as well as electrons
to redistribute themselves. As we mentioned earlier, this results in the
applicability of the MSC-rule instead of the Hund's first rule.

The organization of this paper
is as follows. In section \ref{sec2} we formulate the SSPJM, and show that
in view of the Pauli exclusion principle
the equilibrium bulk density parameter, $r_s$, for nonzero polarization is
somewhat different from that of the spin-compensated system.
Section \ref{sec3} is devoted to the calculational scheme and the application
of the SSPJM to metallic clusters. In section \ref{sec4}, we present the
results of our
calculations of total ground state energy per valence electron, the surface and
curvature energies, the ionization potentials,
and the energy second difference, $\Delta_2(N)$, which is a measure of
relative stability, for different metal
clusters.
Lastly, we explore the condition of the equality of the Fermi energies for the
up- and down-spin bands ($\varepsilon_F^\uparrow=\varepsilon_F^\downarrow$) in
Appendix \ref{app1}.

\section{The Stabilized Spin-Polarized Jellium Model}
\label{sec2}
In this section we generalize the SJM to include a uniform electron system with
nonzero constant spin polarization, $\zeta$. This formulation, in the limit of
$\zeta=0$
reduces to the SJM. The development of the formulation of the SSPJM parallels
the formulation of the SJM. Here, the non-interacting kinetic energy
functional, $T_s$, as
well as the XC-energy functional, $E_{xc}$, depend on the polarization,
$\zeta$,
of the system. But the classical Coulomb interaction of the electrons with
pseudo-ions depends only on their relative distances and is independent of the
spin polarization of the electrons.
In a homogeneous system with nonzero electron polarization,
 the total electron density is the sum over the two spin density components,
i.e., $n=\nup+\ndo$. By defining polarization as $\zeta=(\nup-\ndo)/n$,
each component is expressed in terms of $\zeta$ and $n$

\begin{equation}
\left\{\begin{array}{l}
       \nup=\frac{1}{2}(1+\zeta)n\\
       \ndo=\frac{1}{2}(1-\zeta)n. \end{array}
\right.
\label{eq3}
\end{equation}

As in the SJM, we use the Ashcroft empty core pseudopotential\cite{ash66}
for the interaction between an ion of charge $z$ and an electron at a
relative distance $r$:

\begin{equation}
w(r)=\left\{\begin{array}{ccc}
            -2z/r&,&(r>r_c)\\
            0&,&(r<r_c),
            \end{array}
      \right.
\label{eq4}
\end{equation}
where the core radius, $r_c$, will be fixed by setting the pressure of the
system equal to zero.
The average energy per valence electron in the bulk, with density
$n$ and polarization $\zeta$, is

\begin{equation}
\varepsilon(n,\zeta)=t_s(n,\zeta)+\varepsilon_{xc}(n,\zeta)+\bar
w_R(n,r_c)+\varepsilon_M(n)+\varepsilon_{bs},
\label{eq5}
\end{equation}
where
\begin{equation}
t_s(n,\zeta)=\frac{1}{2}c_k[(1+\zeta)^{5/3}+(1-\zeta)^{5/3}]n^{2/3}
\label{eq6}
\end{equation}
\begin{equation}
\varepsilon_{xc}(n,\zeta)=\frac{1}{2}c_x
[(1+\zeta)^{4/3}+(1-\zeta)^{4/3}]n^{1/3}+\varepsilon_c(n,\zeta)
\label{eq7}
\end{equation}
\begin{equation}
c_k=\frac{3}{5}(3\pi^2)^{2/3}\;\;\;\;c_x=\frac{3}{2}\left(\frac{3}{\pi}
\right)^{1/3}
\label{eq8}
\end{equation}
All equations throughout this paper are expressed in Rydberg atomic units. Here
$t_s$ and $\varepsilon_{xc}$ are the kinetic and exchange-correlation
energies per particle respectively. For $\varepsilon_c$ we use the Perdew-Wang
parametrization \cite{perwan}.
$\bar w_R$
is the average value of the repulsive part of the pseudopotential ($\bar
w_R=4\pi n r_c^2$), and $\varepsilon_M$ is the average Madelung energy of a
collection of point ions embedded in a uniform negative background of density
$n$ ($\varepsilon_M=-9z/5r_0$).
 All nonuniformity of the true electron
density $n(\rr)$ is contained in the band-structure energy $\varepsilon_{bs}$.
$r_0$ being the radius of the Wigner-Seitz sphere, is given by
$r_0=z^{1/3}r_s$. We note that for monovalent metals $z=1$, and if one sets
$z^*=1$ for polyvalent metals, one obtains reasonable agreement with experiment
(see Ref.[\ref{pertran}]). In the latter case, each Wigner-Seitz cell will be
replaced by $z$ smaller cells with volume $4\pi r_s^3/3$ per cell. As in the
SJM, we
assume that $\varepsilon_{bs}$ is negligibly small compared to other terms in
Eq.~(\ref{eq5}).
Since energy, and thereby, pressure in this formalism depend on $\zeta$ as well
as $r_s$, the stabilization of the bulk spin-polarized system with given $r_s$-
and $\zeta$-values, forces the pseudopotential to assume a core radius
appropriate to these values, i.e., $r_c=r_c(r_s,\zeta)$. In order to stabilize
a bulk system with equilibrium
density $\nb$, and constant $\zeta$, we should set the pressure equal to zero:

\begin{equation}
0=P(\nb,\zeta)=-\left(\frac{\partial E}{\partial V}\right)_{N,\zeta}=
\nb^2\left(\frac{\partial}{\partial\nb}\right)_\zeta\varepsilon(\nb,\zeta)=
-\frac{1}{4\pi \bar r_s^2}\left(\frac{\partial}{\partial \bar r_s}\right)_\zeta
\varepsilon(\bar r_s,\zeta).
\label{eq9}
\end{equation}
This equation fixes the core radius, $r_c$, as a function of $\nb$ and $\zeta$.
Using
\begin{equation}
\left(\frac{\partial}{\partial r_s}\right)_\zeta
t_s(r_s,\zeta)=-\frac{2}{r_s}t_s(r_s,\zeta)
\label{eq10}
\end{equation}
and
\begin{equation}
\left(\frac{\partial}{\partial r_s}\right)_\zeta
\varepsilon_x(r_s,\zeta)=-\frac{1}{r_s}\varepsilon_x(r_s,\zeta),
\label{eq11}
\end{equation}
equation (\ref{eq9}) results in
\begin{equation}
2t_s(\bar r_s,\zeta)+\varepsilon_x(\bar r_s,\zeta)-\bar r_s
\left(\frac{\partial}{\partial
\bar r_s
}\right)_\zeta\varepsilon_c(\bar
r_s,\zeta)+\frac{9}{\bar r_s^3}r_c^2+\varepsilon_ M (\bar r_s)=0.
\label{eq13}
\end{equation}
The solution of this equation at equilibrium density, $\nb$, reduces to the
following equation for $r_c$, which will now depend on $\bar r_s$ and $\zeta$,

\begin{equation}
r_c(\bar r_s,\zeta)=\frac{\bar r_s^{3/2}}{3}\left\{
-2t_s(\bar r_s,\zeta)-\varepsilon_x(\bar r_s,\zeta)+\bar r_s
\left(\frac{\partial}{\partial
\bar r_s}\right)_\zeta\varepsilon_c(\bar r_s,\zeta)-\varepsilon_M(
\bar r_s)\right\}^{1/2}.
\label{eq14}
\end{equation}

Because of the $\zeta$-dependence of $r_c$, the difference potential
$\langle\delta
v\rangle_{\rm WS}$ becomes polarization-dependent. Here $\langle\delta
v\rangle_{\rm WS}$ is the average of the difference potential over the
Wigner-Seitz cell and the difference potential, $\delta v$, is defined as the
difference between the pseudopotential of a lattice of ions and the
electrostatic potential of the jellium positive background. As in the SJM
(see Eq. (27) of Ref.
[\ref{pertran}]), at equilibrium density we have through Eq. (\ref{eq9})

\begin{eqnarray}
\langle\delta v\rangle_{\rm WS}&=&\nb\left[\frac{\partial}{\partial n}(\bar
w_R(n) +\varepsilon_M(n))\right]_{n=\nb}\\
&=&-\nb\left[\left(\frac{\partial}{\partial n}\right)_\zeta(t_s(n,\zeta)+
\varepsilon_{xc}(n,\zeta))\right]_{n=\nb},
\label{eq15}
\end{eqnarray}
so that

\begin{eqnarray}
\langle\delta v\rangle_{\rm WS}&=&
  -\nb\left(\frac{\partial}{\partial \nb}\right)_\zeta(t_s(\nb,\zeta)+
+\varepsilon_x(\nb,\zeta))+\frac{\bar r_s}{3}
    \left(\frac{\partial}{\partial
\bar r_s}\right)_\zeta\varepsilon_c(\bar r_s,\zeta)\\
 &=&-\frac{1}{3}\left\{2t_s(\nb,\zeta)+\varepsilon_x(\nb,\zeta)-\bar r_s
\left(\frac{\partial}{\partial
\bar r_s}\right)_\zeta\varepsilon_c(\bar r_s,\zeta)\right\}.
\label{eq16}
\end{eqnarray}

In the above equations, $\nb=3/(4\pi \bar r_s^3)$ is the equilibrium electronic
density of a homogeneous system
which has a nonzero constant polarization. Note that, here, $\nb$ is
 polarization-dependent.
In fact, by increasing the polarization, one increases the
number of the spin-up relative to the spin-down electrons, and therefore, as a
consequence of the Pauli exclusion principle, the total number of Fermi holes
corresponding to the spin-up electrons
is 
increased. This leads to the volume expansion of the system ( It is
a well-known fact that in a molecule, bond lengths depend on spin
configurations). In order to estimate the changes of equilibrium density as a
function of polarization, we put the pressure of the homogeneous spin-polarized
free electron gas equal to zero.
But for a homogeneous spin-polarized free
electron gas with a uniform positive background, we have
\begin{equation}
\varepsilon(r_s,\zeta)=t_s(r_s,\zeta)+\varepsilon_x(r_s,\zeta)+\varepsilon_c(
r_s,\zeta),
\label{eq17}
\end{equation}
where the electrostatic energies have cancelled out. Now, vanishing of the
pressure at equilibrium leads us to
\begin{equation}
0=\frac{1}{4\pi r_s^2}\left\{\frac{2}{r_s}t_s(r_s,\zeta)
+\frac{1}{r_s}\varepsilon_x(r_s,\zeta)-
\left(\frac{\partial}{\partial r_s}\right)_\zeta \varepsilon_c(r_s,\zeta)
\right\},
\label{eq18}
\end{equation}
the solution of which gives the equilibrium $r_s$ as a function of $\zeta$.
At $\zeta=0$, the latter equation yields the well-known paramagnetic value of
4.18, and therefore, in the case of non-zero polarization, it is convenient to
write the equilibrium $r_s$-value as
\begin{equation}
\bar r_s^{\rm EG}(\zeta)=4.18+\Delta r_s^{\rm EG}.
\label{eq18p}
\end{equation}
The increment $\Delta r_s^{\rm EG}$ is used to find a rough estimate for the
equilibrium $r_s$-values of various simple metals through
\begin{equation}
\bar r_s^{\rm X}(\zeta)=\bar r_s^{\rm X}(0)+\Delta r_s^{\rm EG},
\label{eq18z}
\end{equation}
where the superscript X refers to a given simple metal.
In
applying this result to different simple metals, we assume that this
increment is independent of the value of $\bar r_s(\zeta=0)$, and simply
depends on the value of
$\zeta$.
This is the simplest assumption we have thought of. Other forms of
$\bar r_s^{\rm
X}(\zeta)$ is possible. For instance, one could use a low degree polynomial of
$\zeta$ with coefficients depending on the type of metal.
 For Al, Ga, Li, Na, K, and Cs, $\bar r_s(0)$
are taken to be 2.07, 2.19, 3.28, 3.99, 4.96, 5.63 respectively.
The increment of equilibrium
density $\nb$
due to increasing $\zeta$, affects the value of the core radius of the
pseudopotential and causes it to increase monotonically as a function of
 $\zeta$.
 Also the value of
$\langle\delta v \rangle_{\rm WS}$ will depend on $\zeta$.
Fig.~\ref{fig1} shows the behavior of $\langle\delta v\rangle_{\rm WS}$ as a
function of polarization, $\zeta$, for the six metallic systems considered in
here. One notes
that for metals with the value of $\bar r_s(0)$  less than 4.18, the increase
in  $\zeta$ weakens the effective potential relative to its value at
$\zeta=0$, but for metals with $\bar r_s(0)>4.18$, the depth of the potential
increases. Once the values of $\langle\delta v\rangle_{\rm WS}$ and $r_c$ as
a function of $\bar r_s$ and $\zeta$ are found, the equation (23) of
Ref.~[\ref{pertran}] can be generalized to
\begin{eqnarray}
E_{\rm SSPJM}[\nup,\ndo,n_+]&=&
E_{\rm JM}[\nup,\ndo,n_+]+(\varepsilon_M(\nb)+\bar w_R(\nb,
\zeta))\int d\rr\;n_+(\rr) \nonumber \\
  &&+\langle\delta v\rangle_{\rm WS}(\nb,\zeta)\int
d\rr\;\Theta(\rr)[n(\rr)-n_+( \rr)],
\label{eq19}
\end{eqnarray}
where
\begin{eqnarray}
E_{\rm JM}[\nup,\ndo,n_+]&=&T_s[\nup,\ndo]+E_{xc}[\nup,\ndo] \nonumber\\
&&+\frac{1}{2}\int d\rr\;\phi([n,n_+];\rr)[n(\rr)-n_+(\rr)],
\label{eq20}
\end{eqnarray}
and
\begin{equation}
\phi([n,n_+];\rr)=\int d\rrp\;\frac{[n(\rrp)-n_+(\rrp)]}{\mid\rr-\rrp\mid}.
\label{eq21}
\end{equation}
$\Theta(\rr)$ has the value of unity inside the system and zero outside.
By taking the variational derivative of $E_{\rm SSPJM}$ with respect to
spin densities $\nup,\ndo$, one finds the Kohn-Sham (KS) effective potentials
\begin{eqnarray}
v_{eff}^\sigma(\rr,\zeta)&=&\frac{\delta}{\delta n_\sigma(\rr)}(E_{\rm SSPJM}
-T_s)\\
&=&\phi([n,n_+];\rr)+v_{xc}^\sigma(\rr)+\Theta(\rr)\langle\delta
v\rangle_{\rm WS} (\nb,\zeta),
\label{eq22}
\end{eqnarray}
where $\sigma=\uparrow,\downarrow$. The distinction between the
$v_{eff}^\sigma$ for the two schemes of the SSPJM and the SJM is rooted in the
$\zeta$-dependence of the quantities $n_+$ and $\langle\delta v\rangle_{\rm
WS}$ for the SSPJM. The forms
of $v_{xc}^\sigma (\rr)$ are the same in both cases.
By solving the KS-equations
\begin{equation}
(\nabla^2+v_{eff}^\sigma(\rr))\phi_i^\sigma(\rr)=\varepsilon_i^\sigma
\phi_i^\sigma(\rr)\;\;\;;\;\;\;\;\sigma=\uparrow,\downarrow
\label{eq23}
\end{equation}
\begin{equation}
n(\rr)=\sum_{\sigma=\uparrow,\downarrow}n_\sigma(\rr),
\label{eq24}
\end{equation}
\begin{equation}
n_\sigma(\rr)=\sum_{i(occ)}\mid\phi_i^\sigma(\rr)\mid^2,
\label{eq25}
\end{equation}
and finding the self-consistent values for $\varepsilon_i^\sigma$ and
$\phi_i^\sigma$, one obtains the total energy.
In the next section we apply this model to simple metal clusters.

\section{Calculational Scheme}
\label{sec3}
In the spherical-SSPJM with sharp boundary, the positive background
charge density is constant and equals to $\nb(\zeta)$ inside the sphere  of
radius $R(\zeta)=N^{1/3}\bar r_s(\zeta)$ and vanishes outside.
In the case of dif-SSPJM, since there exists no sharp boundary for the
jellium, we have taken the effective boundary at
$R(\zeta)=N^{1/3}\bar r_s(\zeta)$.
 Here, for metal clusters, we take the value of $\zeta$ as
\begin{equation}
\zeta=(N_\uparrow-N_\downarrow)/N\;\;\;\;\;;\;\;\;
N=N_\uparrow+N_\downarrow,
\label{eq29}
\end{equation}
where $N_\uparrow$ and $N_\downarrow$ are the total numbers of valence
electrons with spins up and down respectively.
 One could instead take $\zeta$  as the following average:
 \begin{equation}
\bar\zeta=\frac{1}{\Omega}\int_0^{r_1}4\pi r^2\;dr
\frac{[\nup(r)-\ndo(r)]}{n(r)} \;\;\;\;\;\;\\;\;\;\;\Omega=\frac{4\pi}{3}r_1^3,
\label{eq30}
\end{equation}
 in which $r_1$  is the radius at which the density components fall to the
value
of, say 1 percent of their peak values. In the homogeneous case, the equation
for
$\bar\zeta$ reduces to our assumed relation. The densities in the integrand of
$\bar\zeta$ are the
self-consistent values obtained by solving the KS-equations. It
turns
out that these values of $\zeta$  and $\bar\zeta$ are very close to each
other and so
we use the simpler one. The effective potentials in the  KS-equation will be
\begin{equation}
v_{eff}^\sigma(\rr;[\bar r_s,\zeta])=v_b(\rr;[\bar r_s])+v_{\rm
H}(\rr;[\bar r_s])+ v_{xc}^\sigma(\rr;[\bar r_s,\zeta])+
\Theta(\rr)\langle\delta v\rangle_{\rm WS}(\bar r_s,\zeta),
\label{eq31}
\end{equation}
where in the case of sphere with sharp boundary,
\begin{equation}
v_b(r)=\left\{\begin{array}{lcc}
                  -(N/R)[3-(r/R)^2]&;&r\le R\\
                  -2N/r&;&r>R. \end{array}
             \right.
\label{eq32}
\end{equation}
In the above equation, $v_b$ is the potential energy of interaction between
an electron and the positive charge background which depends on $\zeta$ via
 $r_s(\zeta)$. Also we have
\begin{equation}
v_{\rm H}(\rr)=2\int\frac{n(\rrp)}{\mid\rr-\rrp\mid}\;d\rrp,
\label{eq33}
\end{equation}
\begin{equation}
v_{xc}^\sigma(\rr)=\frac{\delta E_{xc}[\nup,\ndo]}{\delta n_\sigma(\rr)}.
\label{eq34}
\end{equation}

A spherical jellium with a finite surface thickness (diffuse-jellium)
 \cite{alon}, is defined by

\begin{equation}
n_+(r)=\left\{\begin{array}{lcc}
              \nb\{1-(R+t)e^{-R/t}[\sinh(r/t)]/r\}&;&r\le R\\
              \nb\{1-((R+t)/2R)(1-e^{-2R/t})\}R e^{(R-r)/t}/r&;&
               r>R, \end{array}
        \right.
\label{eq35}
\end{equation}
 where $R=N^{1/3}r_s$, and $t$ is a parameter related to the surface
thickness
(for other forms of diffuse-jellium see Ref.[\ref{brack93}]).
In our numerical calculations with diffuse-jellium for Na-clusters, we have
chosen $t=1.0$ both for neutral and singly ionized clusters.
In the
case of dif-SSPJM
the potential energy of an electron due to the background will be
\begin{equation}
v_b^{dif}(r)=\left\{\begin{array}{lcc}
                       (4\pi/3)\nb (3R^2-r^2-6t^2)-8\pi\nb t^2(R+t)
                       e^{-R/t}[\sinh(r/t)]/r&;&r\le R\\
                       -2N/r+4\pi\nb t^2[(R+t)+(R-t)e^{2R/t}]
                       e^{-(R+r)/t}/r&;&r>R. \end{array}
                 \right.
\label{eq36}
\end{equation}
Here, $R$ is the effective radius of the jellium sphere, i.e., $R=N^{1/3}\bar
r_s(\zeta)$.

\section{Results and Discussions}
\label{sec4}
In this section we discuss the calculated results for metal
clusters (Al, Ga, Li, Na, K, Cs) of different sizes $(2\le N\le 42)$. After
computing the KS-orbitals for spin-up and spin-down components, and finding the
corresponding eigenvalues,
 the total energy of an $N$-electron cluster, $E(N)$,
 were calculated for both sharp- and diffuse-jellium spheres. For the sake of
comparison, we have also repeated the calculations based on the JM and the
SJM using the local spin density approximation (LSDA).
 For a given
$N$-electron cluster, we have considered the various combinations
of $N_\uparrow$ and $N_\downarrow$ values while keeping the total number of
valence
electrons, $N=N_\uparrow +N_\downarrow$, fixed.

Our calculations based on Eq.~(\ref{eq19}), show that the total
energy of the system decreases as its polarization decreases, i.~e., the system
moves towards spin-compensated configurations. In fact, the total energy
minimization is accomplished when electron spin compensation is maximum.
Namely, for clusters with even number of electrons, the energy minimization
gives rise to complete compensation ($N_\uparrow=N_\downarrow$),
and for
clusters with odd number of electrons, only one electron remains uncompensated
($N_\uparrow-N_\downarrow=1$).

In order to show the effect of MSC-rule, we have plotted
 $\Delta$MSC$=E_{\rm flipped}-E_{\rm MSC}$, as a function of $N$ in 
Fig.~\ref{fig2} for Na- and Al-clusters. Here, $E_{\rm MSC}$ is the total
energy of a cluster assuming MSC-configuration, and $E_{\rm flipped}$ is the
 total energy of the same cluster, but with a configuration involving only
one spin-flip in the outer shells relative to the MSC-configuration. The
spin-flip is to be consistent with the Pauli exclusion principle.
In fact, we have calculated $\Delta$MSC for Al$_n$, Ga$_n$, Li$_n$, Na$_n$,
 K$_n$, and Cs$_n$-clusters with $2\le nz\le 42$,
 and found that it is always positive except for Al$_{12}$, Al$_{13}^+$,
 Al$_{14}$, and Ga$_{12}$. Thus, we conclude that MSC-rule is at work for
 nearly all clusters considered in here. 

It is worth noting that in the JM and the SJM calculational schemes the minimum
 energy corresponds to configurations with maximum spin polarization in the 
outermost shells; i.e., Hund's first rule is at work. For example, in a
13-valence electron cluster, the JM and the SJM result in the minimum energy
 configuration in which all 5 electrons in the outermost shell ($l=2$) are
 in parallel spin-up state ($N_\uparrow -N_\downarrow =5$), while in the SSPJM
the energy is minimized when $N_\uparrow -N_\downarrow =1$.

The reasons for different behaviors resulting from the use of the JM and the
SJM or of the SSPJM and the dif-SSPJM, namely the applicability of the Hund's
first rule or the MSC-rule lies in the unnecessary constraint of rigid jellium
 background assumed in earlier JMs. Note that, in these models one fixes
the $r_s$-value for all cluster sizes with arbitrary spin configurations.  
 Lifting such a contraint
is consistent with the well-known fact that molecular bond-lengths depend on
spin polarization. Moreover,
from molecular
dynamics calculations one can infer the alternating variation of the nearest
neighbor distances
as a function of cluster sizes\cite{roth}, which supports our idea of
allowing the alternating volume expansion of the jellium background positive
charge distribution. In cases where the outermost shell is closed 
 ($N=$ 2, 8, 18, 20, 34, 40, $\ldots$), all the four schemes
of the JM, SJM, SSPJM, and dif-SSPJM predict the same spin configurations
 and the radius of the jellium sphere is the same for the first three schemes,
but differs in the dif-SSPJM. Also, the total energy values are the same both
in the SSPJM and the SJM. In cases where the outermost shell contains only one
electron or lacks one electron to have a closed shell ($N=$ 1, 3, 7, 9, 17,
 19, 21, 33, 35, 39, 41, $\ldots$), the above-mentioned four schemes predict
 the same 
spin configurations, but the energies are all different. Here, the difference 
in the SJM- and the SSPJM-values arise from different jellium radii in the 
two schemes for a given cluster size. In the above two special cases, the 
Hund's first rule and the MSC-rule are identical.

When the cluster size remains fixed while the
 polarization changes, as was assumed in the case of the JM and the SJM, the
only way the system
could reduce its total energy was to redistribute its electrons further apart.
 One
finds the situation in the JM and the SJM to be very similar to that of atoms
in which the external potential of the nucleus is kept fixed for different spin
configurations.
As in the case of atoms, due to the Pauli exclusion principle, parallel spin
electrons are required to stay apart, i.~e., further apart than when they
assume a spin-compensated configuration.
In the SSPJM considered in this paper,
 the relative positive charge
background radii, $R(N,\zeta)/R(N,0)$, of clusters are allowed to expand with
increasing polarization. In a sense, here, the ionic motions are simulated
through such an expansion. Now, because of the above-mentioned freedom, as soon
as electrons try to take advantage of the
Pauli exclusion principle and begin to spill out of the cluster, the positive
charge background will try to follow them. In other words, the freedom of
cluster size expansion renders the application of Hund's first
rule unnecessary, and the SSPJM chooses to be in a spin-compensated
configuration.
 In short, contrary to the case of the spherical-JM and the spherical-SJM,
which are governed by the Hund's first rule, here the MSC-rule is in
effect, and
it is this MSC effect which results in the changes in the level separations
near the Fermi level for neutral and ionized clusters,  and thereby, gives rise
to the
well-known odd-even alternations in IPs of alkali metal clusters.
 One notes that our viewpoints about
the changes in level separations, and those that attribute this effect
to  nonspherical shape deformations
 refer to two complementary effects.
That is, in our case, the shape has remained spherical whereas the volume is
allowed to
change, but in the case of deformed-JMs, the volume is fixed and
the shape is allowed to change. In a combined effect, one allows, at the same
time, the volume of a cluster to change and its shape to deform.

 The size of jellium sphere in the SSPJM for neutral and singly ionized
clusters are
different. This should be contrasted with the cases of the JM and the SJM
in which the sizes for the neutral and ionized clusters are assumed to be
the same. Our self-consistent calculations show that the polarization $\zeta$,
for metal clusters in their minimum energy
configuration as a function of total number of valence electrons, $N$,
satisfies (with exceptions for Al$_{12}$, Al$_{13}^+$, Al$_{14}$, Ga$_{12}$)
the following equation:

\begin{equation}
\zeta(N)=\frac{[1-(-1)^N]}{2N}=\left\{\begin{array}{ccc}
                                      0&;&N\;\;{\rm even}\\
                                      1/N&;&N\;\;{\rm odd.}\end{array}
                               \right.
\label{eq37}
\end{equation}

Using the above equation for polarization, we plot the equilibrium $r_s$-value
for different Na-clusters in Fig.~\ref{fig3}. As seen from the figure, for
 an Na-cluster with even number of atoms, $r_s$  equals
3.99 and with odd number of atoms the envelope is a decreasing function of $N$.
According to Eq.~(\ref{eq18z}), Fig.~\ref{fig3} for other metals remains the
same but shifted according to the value of $r_s(0)$. As is expected, in the
 limit of large-$N$  clusters, 
addition or removal of  an electron does not change the configurations of
 all other ions. Comparing Fig.~\ref{fig3} with Fig. 15(a) of
Ref. [\ref{roth}] for average nearest neighbor distances,
one notes the same staggering effect.

Fig.~\ref{fig4}a and Fig.~\ref{fig4}b show the values of the nonbulk binding
energies, $E/N-\alpha_V$, for Na- and Al-clusters using the JM, the SJM,
and the SSPJM along with the LSDA. Here, $\alpha_V$ represents the bulk binding
 energy.
The calculated energies based on the JM, the SJM, and the SSPJM for Na-clusters,
aside from the details, are nearly close and positive as expected. This is
because the sodium $r_s$-value (3.99) is close to the zero-pressure jellium
$r_s$-value (4.18). However, in the case of Al-clusters, the energies based
 on the SJM and the SSPJM remain near each other and are positive, but the
 energies based on the JM is considerably away from them and at times assume
negative values (see Fig.~\ref{fig4}b). The reason is that the JM at high
 densities fails (note that the $r_s$-value of Al is 2.07) and leads to 
mechanical instability. We have also calculated the nonbulk binding energies
of other metal clusters (Ga, Li, K, Cs) and see that as $r_s$ decreases,
the energies based on the JM moves away from the results based on the SJM
and the SSPJM. By looking at Fig.~\ref{fig1} we note that the $\langle\delta v
\rangle_{\rm WS}$ contribution to the total energies are positive in the case
of Cs and K (for which $r_s>4.18$); and negative in the case of Al, Ga, Li,
and Na (for which $r_s<4.18$). Thus, in the case of K- and Cs-clusters we
 should
expect that the JM energies using the LSDA lie above the SJM- and the
 SSPJM-energies, and in 
the case of Na-, Li-, Ga-, and Al-clusters we should expect it to lie below 
them. 
Our calculations of the nonbulk energies of Al-, Ga-, Li-, Na-, K-, Cs-clusters
 confirm
these conclusions. 

In Fig.~\ref{fig5} we show the plot of $\Delta_2(N)=E(N+1)+E(N-1)-2E(N)$
which determines the relative stability of different Na-clusters. We have also
 calculated $ \Delta_2(N)$ for other metal clusters mentiond in this paper.
In the plot using the SSPJM, the incorrect peaks predicted by the spherical-SJM
at $N=$5, 13, 27, 37 have disappeared and the clusters with  $N=$8, 18, 20, 40
 are predicted to be more stable. These results are consistent with the 
fine-structure observed experimantally in the abundance curve \cite{deheer93}.
Similar observations can be made from the plots of $\Delta_2(N)$ for other
metals considered in here. The overall agreement between our results and
 experimental data is good.

Next, we have calculated the surface and curvature energies of Al-, Ga-, Li-, 
Na-, K-, Cs-clusters. According to the liquid drop model
\cite{brack89,engel91,yano95}, one may
 write the total energy of a finite quantal system in the form of 

\begin{equation}
E=\overline{E}+\delta E
\label{eq38}
\end{equation} 
where $\delta E$ is shell correction and $\overline E$ is the smooth part of
 the total energy which is written as
a sum of volume, surface, and curvature contributions. In the
case of spherical geometry, $\overline E$ reduces to the following
 parametrized equation
as a function of the number of valence electrons in a neutral cluster 

\begin{equation}
\overline{E}(N)=\alpha_V N + \alpha_S N^{2/3}+\alpha_C N^{1/3}.
\label{eq39}
\end{equation}
Here, $\alpha_V$ is the total energy per electron in the bulk.  
Its absolute values for the metals Al, Ga, Li, Na, K, Cs using the SJM or
the SSPJM are respectively 10.57, 10.14, 7.37, 6.26, 5.18, 4.64 in units of
 electron-volts using the $r_s$-values mentioned earlier.
 The surface energy, $\sigma$, and the curvature energy, 
$A_C$, are related to the parameters $\alpha_S$ and $\alpha_C$ through
the relations:

\begin{equation}
\sigma=\frac{1}{4\pi r_s^2}\alpha_S,\;\;\;\;\;A_C=\frac{1}{4\pi r_s}\alpha_C.
\label{eq40}
\end{equation} 

The parameters $\alpha_S$ and $\alpha_C$ are obtained by a least-square fit
of our self-consistent total energies to Eq.~(\ref{eq39}).

Fig.~\ref{fig6}a shows the surface energies of different metals as a function
of their $r_s$-values for the three schemes using the SSPJM, the SJM, and the
 JM, and compared with the results obtained by others (BULK) \cite{fiol92}.
 The surface
energies of the SSPJM and the SJM remain positive for the metals considered
here, but it becomes negative in the case of the JM (as mentioned in the
 introduction)
 for high electron density metals (Al, Ga). 

Fig.~\ref{fig6}b shows the curvature energies of various metals as a function
 of their $r_s$-values. The SSPJM results in a higher curvature energies
than the other two schemes using the SJM and the JM. The results of bulk
calculations \cite{fiol92} are also shown. 

Finally, in Figs.~\ref{fig7}a-f, we have compared the ionization energies of
metal clusters for different schemes as well as with the experimental values.
The ionization energy of an $N$-valence electron cluster is defined as the
 difference
between the total ground state energy of that system with $N$ and ($N-1$)
 valence electrons.
The experimental values are taken from Ref. [\ref{kenneth}] for aluminum,
Ref. [\ref{deheer93}] for lithium, sodium, and potassium.
When we calculate the IPs by means of the SSPJM and the SJM, we see that the
odd-even alternations in alkali metal clusters, not present in the results of
the JM or the SJM
 results, show themselves up in the SSPJM calculations.
In the SJM using the
LSDA, kinks appear only at half-filled and closed shells, whereas in both
the SSPJM
and the dif-SSPJM, there exists a peak at the middle of each pair of adjacent
odd numbers.
These peaks correspond to clusters with even number of atoms, in agreement with
experimental data. In dif-SSPJM, the ionization energies of Na-clusters are
 lowered so
that the values near the closed shells (where the nonsphericity becomes less
important) are in good agreement with experiment than the corresponding values
 of the other two models. At the end,
because of the assumed spherical geometry, the pronounced shell effects are
still present in the IPs when we go from one closed shell cluster to a cluster
with one more electron and the overall saw-toothed behavior of IPs remains.
In other metals the agreement between theory and experiment is poor for all
the three schemes.
We think our results for Na-clusters will improve if dif-SSPJM is used along
 with spheroidal or ellipsoidal geometries.

\section{SUMMARY AND CONCLUDING REMARKS}
\label{conclusion}

In this paper, we have generalized the SJM to the
spin-polarized case by allowing the volume of the spherical positive charge
background to change and have calculated the energetics of
metal clusters (Al, Ga, Li, Na, K, Cs).
Our self-consistent calculations show that for spherical geometries, the
minimum-energy of a metal cluster is obtained when the electronic spin
compensation is
maximum. That is, in contrast to the spherical-JM and the spherical-SJM which
are governed by the Hund's first
rule, here the MSC-rule is in effect. We have discussed in
detail
that the situations in the JM and the SJM are similar to that in atoms. In both
cases, the
external potential -- being due to the positive charge background
or due to the nuclear charge -- are fixed and therefore, in both cases
the  Hund's first rule is applicable. However, in the
case of the SSPJM, because of the extra degrees of freedom, namely the
expansion of the positive charge background, the system assumes maximum
spin-compensated configurations.
 This MSC-rule results in the alternations of the average
distance between the nearest neighbors, and thereby, in the alternations of the
IPs.
Moreover, application of the dif-SSPJM for alkali metal clusters brings the
IP-values closer to the experimental data. Finally, we believe that if the 
SSPJM is used in conjunction with nonspherical shape deformations, better
 agreements between theory and experiment will result.

\newpage
\vskip 1truecm 
\begin{center}
{\bf ACKNOWLEDGEMENT}
\end{center}
M.P. is indebted to John P. Perdew for his helpful comments, and also Erick Koch
for the discussion. N.N. would like to thank N. H. March and U.
R\"othlisberger for their helpful discussions.
The authors would like to acknowledge the International Center for Theoretical
Physics, Trieste, Italy, for their warm hospitality.

\appendix\section{STABILITY CONDITIONS IN A SPIN-POLARIZED INFINITE
JELLIUM }
\label{app1}
In an unpolarized infinite jellium ($\zeta=0$), the stability of the system is
obtained via $\partial \varepsilon/ \partial r_s \mid_{\zeta =0}=0$. In other
words,
in this case, it suffices to establish mechanical stability. This is done by
employing the Ashcroft empty-core pseudopotential whose core radius is fixed by
setting the pressure equal to zero. However, for a spin-polarized infinite
jellium, the total energy per electron, $\varepsilon$, depends on $n$ and
$\zeta$. Thus
\begin{equation}
d\varepsilon=\left(\frac{\partial \varepsilon}{\partial n}\right)_\zeta dn +
   \left(\frac{\partial \varepsilon}{\partial \zeta}\right)_n d\zeta.
\label{appeq1}
\end{equation}
At equilibrium, the following two equations must be satisfied:
\begin{equation}
\left(\frac{\partial \varepsilon}{\partial n}\right)_\zeta=0
\label{appeq2}
\end{equation}
\begin{equation}
\left(\frac{\partial \varepsilon}{\partial \zeta}\right)_n=0.
\label{appeq3}
\end{equation}
Eq.~(\ref{appeq2}) establishes the mechanical stability and Eq.~(\ref{appeq3})
sets the Fermi energy of the up- and down-spin bands equal, i.~e.,
$\varepsilon_F^{\uparrow}=\varepsilon_F^{\downarrow}$.  The latter point is
proved below. $\varepsilon_F^{\uparrow}$ is the highest occupied spin-up
KS-level and $\varepsilon_F^{\downarrow}$ is the highest occupied spin-down
KS-level.

\noindent{\bf proof:}

Following the work of Russier, Salahub, and Mijoule \cite{russier}, we can
write
\begin{equation}
\left(\frac{\partial \varepsilon}{\partial \zeta}\right)_n=
\frac{\partial \varepsilon}{\partial n_\uparrow}\left(\frac{\partial
n_\uparrow}{\partial \zeta}\right)_n +
\frac{\partial \varepsilon}{\partial n_\downarrow}\left(\frac{\partial
n_\downarrow}{\partial \zeta}\right)_n,
\label{appeq4}
\end{equation}
so that using Eq.~(\ref{eq3}) reduces to
\begin{equation}
\left(\frac{\partial \varepsilon}{\partial \zeta}\right)_n=
\frac{1}{2}n\left(\frac{\partial \varepsilon}{\partial
n_\uparrow}-
\frac{\partial \varepsilon}{\partial n_\downarrow}
\right).
\label{appeq5}
\end{equation}
Now, using the variational principle for the ground state energy of a
homogeneous system subject to the two constrains
\begin{equation}
\int n_\sigma d\rr=N_\sigma,\;\;\;\;\;\;\;\;\;\sigma=\uparrow,\;\downarrow,
\label{appeq6}
\end{equation}
we have
\begin{equation}
\delta\left\{\int(\nup+\ndo)\varepsilon(\nup,\ndo) d\rr -\mu_\uparrow\int
\nup d\rr -\mu_\downarrow\int\ndo d\rr \right\}=0.
\label{appeq7}
\end{equation}
This equation results in the following Euler equations:
\begin{equation}
\varepsilon+n\frac{\partial\varepsilon}{\partial\nup}-\mu_\uparrow=0,
\label{appeq8}
\end{equation}
\begin{equation}
\varepsilon+n\frac{\partial\varepsilon}{\partial\ndo}-\mu_\downarrow=0.
\label{appeq9}
\end{equation}
Subtracting Eq.~(\ref{appeq9}) from Eq.~(\ref{appeq8}) and dividing both
sides by $n$, one obtains
\begin{equation}
\left(\frac{\partial \varepsilon}{\partial n_\uparrow}-
\frac{\partial \varepsilon}{\partial n_\downarrow} \right) =
\frac{1}{n}(\mu_\uparrow -\mu_\downarrow).
\label{appeq10}
\end{equation}
Substituting Eq.~(\ref{appeq10}) into Eq.(\ref{appeq5}), one obtains
\begin{equation}
\left(\frac{\partial\varepsilon}{\partial\zeta}\right)_n=\frac{1}{2}
 (\mu_\uparrow -\mu_\downarrow).
\label{appeq11}
\end{equation}
Eqs.~(\ref{appeq11}) and (\ref{appeq3}) in conjunction with Koopmans' theorem
\cite{koop}
entails the equality of the two Fermi levels for an infinite homogeneous
spin-polarized jellium.

However, in this paper we have calculated the total energy of finite clusters.
In such systems, because of the discrete nature of the energy eigenvalues, the
equality of the spin-up and spin-down Fermi energies does not hold any more.
Thus, the stabilization will be established through Eq. (\ref{appeq2}) and when
$\partial E/\partial\zeta\mid_{r_s}$ changes sign. In practice, for a given
$\zeta$, we have first calculated the core radius entering the Ashcroft
pseudopotential via Eq. (\ref{eq14}) and used it as an input for the total
energy calculation of the cluster. We then, varied $\zeta$, i.~e., the
cluster's
spin configuration, till total energy minimization was attained. Although,
the
procedure employed here is a consistent choice, but one could start with a
two parameter
pseudopotential and apply Eq. (\ref{appeq2}) and Eq. (\ref{appeq3}) to the
stabilized spin-polarized infinite jellium. In this way, one obtains the
dependence
of the two mentioned parameters on $\zeta$ and $r_s$ and repeats the clusters'
total energy calculations again. Work in this direction is in progress.

\newpage
\begin{figure}
\noindent
\caption{
Equilibrium potential difference in Rydbergs versus polarization for Al, Ga, 
Li, Na, K, and Cs metals.}
\label{fig1}
\end{figure}

\begin{figure}
\noindent
\caption{The behavior of $\Delta$MSC as a function of number of valence 
electrons, $N$, in units of eV. Both neutral and singly ionized metal 
clusters (Al and Na) are presented. $\Delta$MSC is the difference between
the energy of the maximum spin compensated electronic configuration from
the energy of a different configuration in which one electron is flipped. In 
the case of Na-clusters all $\Delta$MSCs are positive and in the case of
Al-clusters, except for Al$_{12}$, Al$_{13}^+$, Al$_{14}$, all $\Delta$MSCs
are again positive. }
\label{fig2}
\end{figure}

\begin{figure}
\noindent
\caption{Equilibrium Wigner-Seitz radius in atomic units as a function of
 number of atoms in an Na-cluster.}
\label{fig3}
\end{figure}

\begin{figure}
\noindent
\caption{The nonbulk total energies per atom of (a) Na- and (b) Al-clusters
as a function of number of valence electrons in the cluster. The symbols,
 solid squares,
diamonds, and triangles correspond to the SSPJM, the SJM, and the JM 
respectively. For Al, the physical points (i.e., multiples of 3)
 are specified and for Na, the large
square symbol corresponds to dif-SSPJM.}
\label{fig4}
\end{figure}

\begin{figure}
\noindent
\caption{The second difference of total energies of Na-clusters in units
of eV as a function of number of atoms in a cluster. The symbols, 
solid squares, triangles, and diamonds correspond to SJM, JM, and dif-SSPJM 
respectively. }
\label{fig5}
\end{figure}

\begin{figure}
\noindent
\caption{(a) Surface energies of metals, $\sigma$, as a function of their
$r_s$-values, in units of erg/cm$^2$; (b) Curvature energies of metals,
$A_c$, as a function of their $r_s$-values, in units of mili-hartree/bohr.
The symbols, crosses, solid squares, triangles, and upside-down triangles
correspond to the SSPJM, SJM, JM, and bulk calculations (BULK)
(see Ref. [\protect\ref{fiol92}]). 
The negative surface energies of Al and Ga are due to the instability of
 the JM. }
\label{fig6}
\end{figure}

\begin{figure}
\noindent
\caption{The ionization energies in units of eV as a function of the number
of atoms in a cluster for (a) Al, (b) Ga, (c) Li, (d) Na, (e) K, (f) Cs.
The large solid squares represent the experimental data. The small solid
squares, the diamonds, the triangles, and the upside-down triangles,
 respectively
correspond to calculational schemes based on the SSPJM, the SJM, the JM, and
the dif-SSPJM. By moving towards closed shell electronic structures,
 the agreement between the results of the dif-SSPJM and experimental data
improves. }
\label{fig7}
\end{figure}

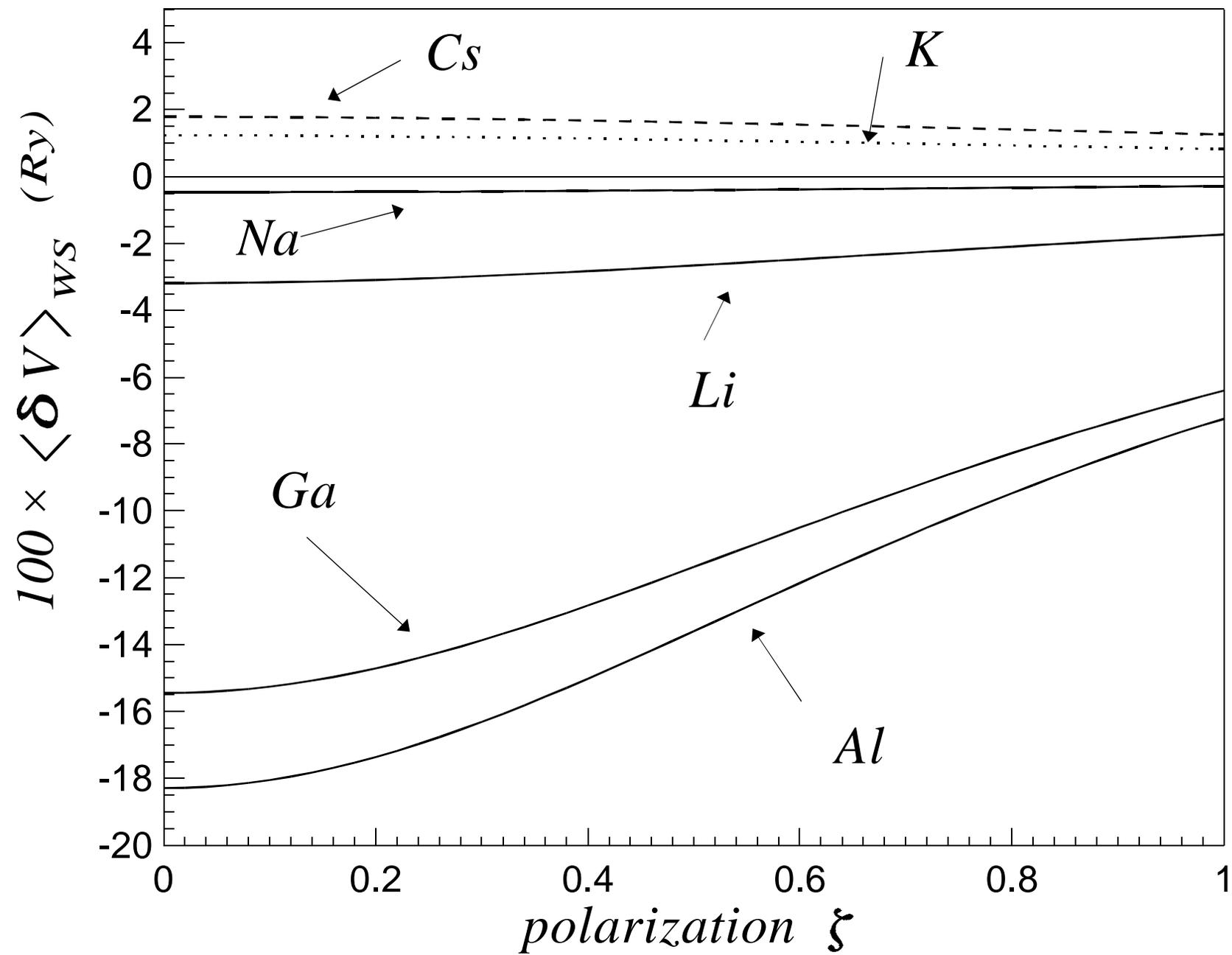
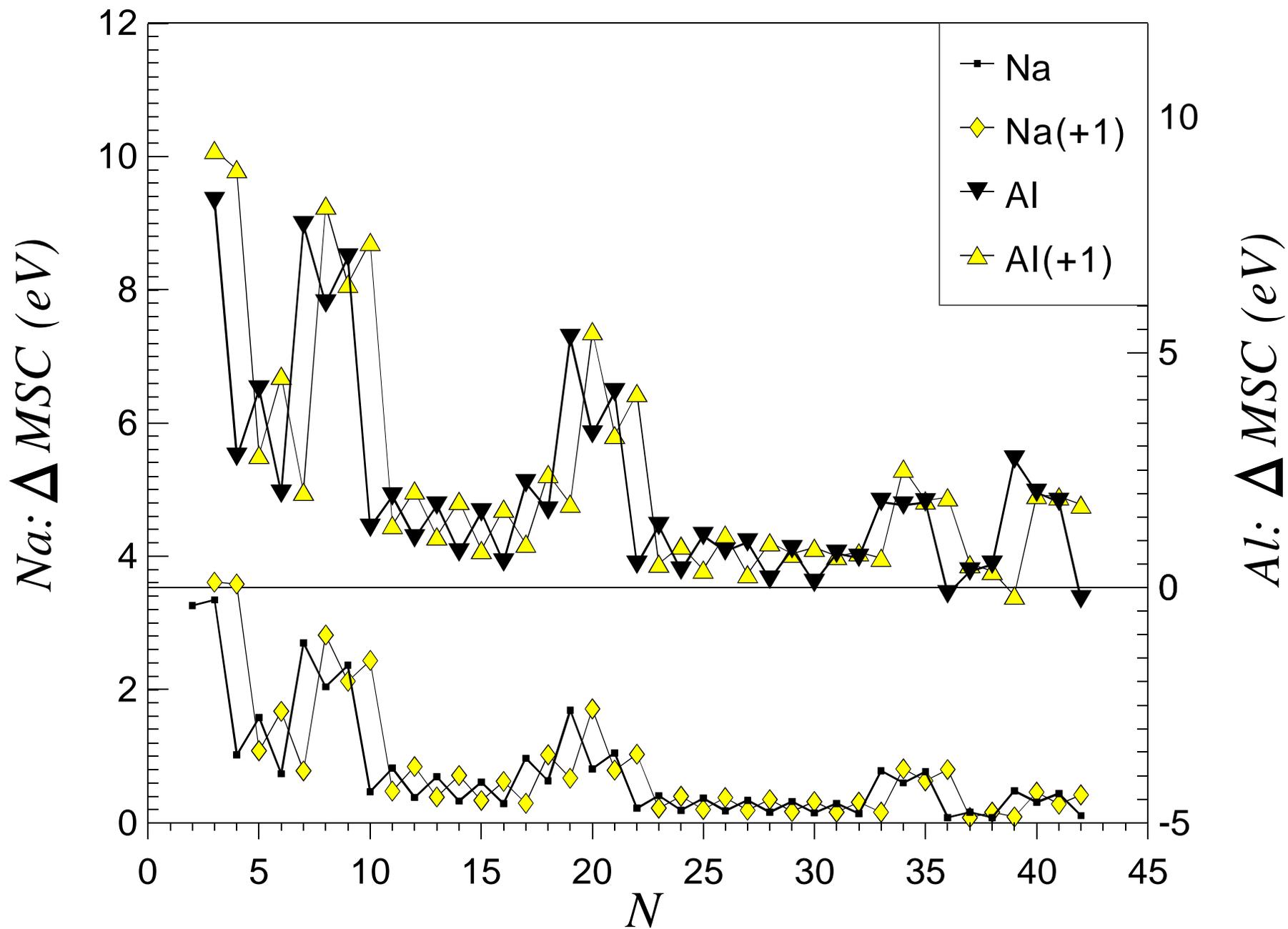
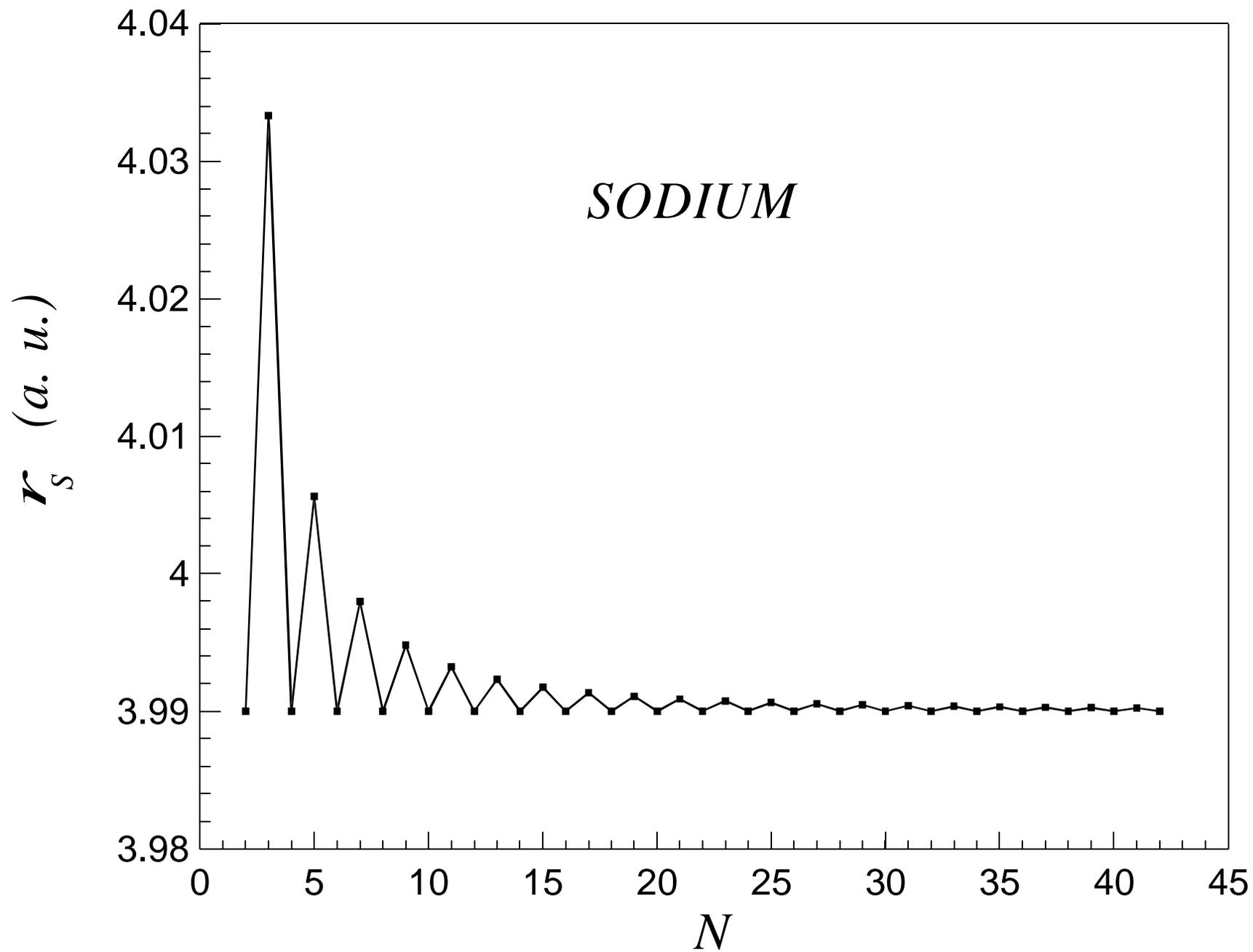
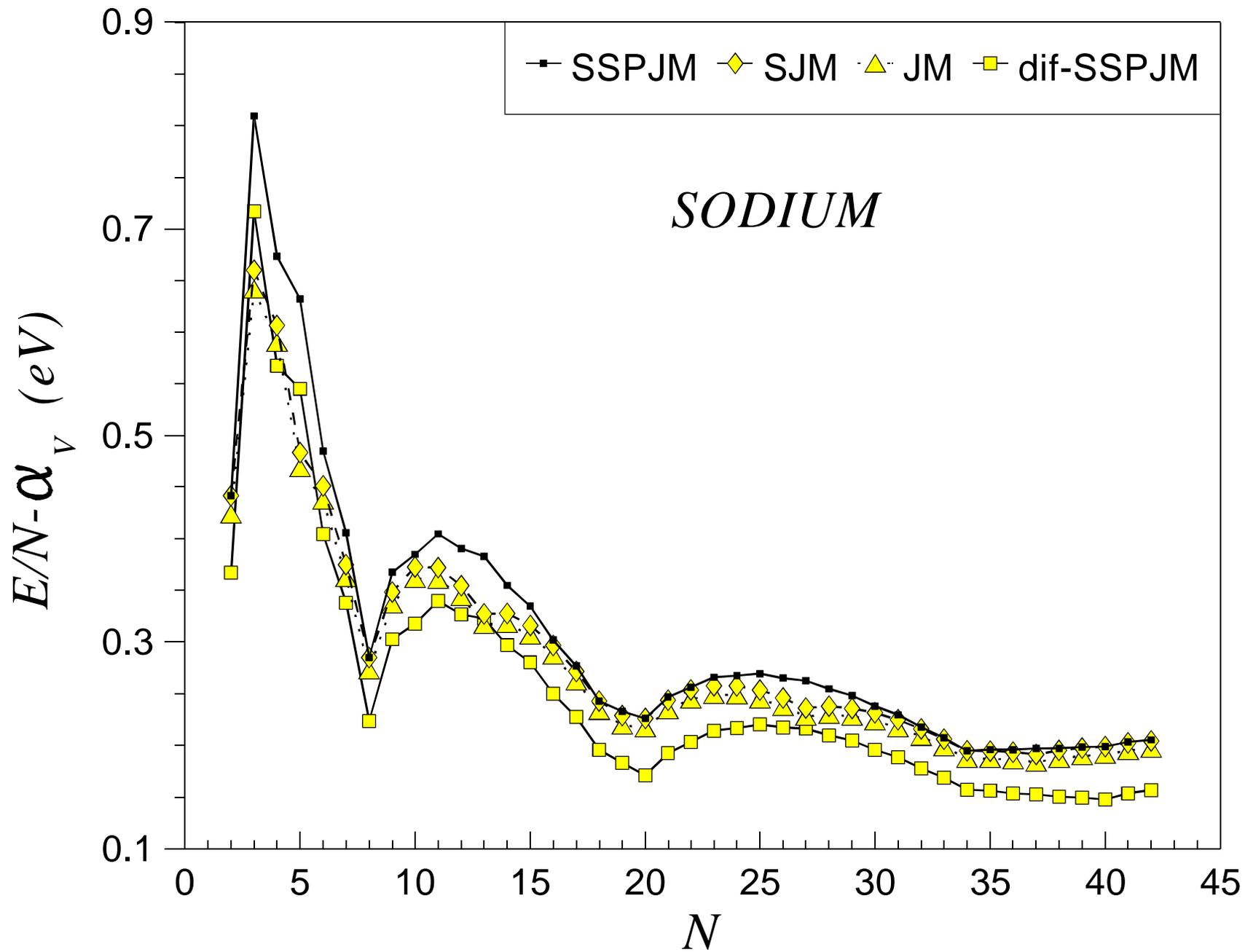
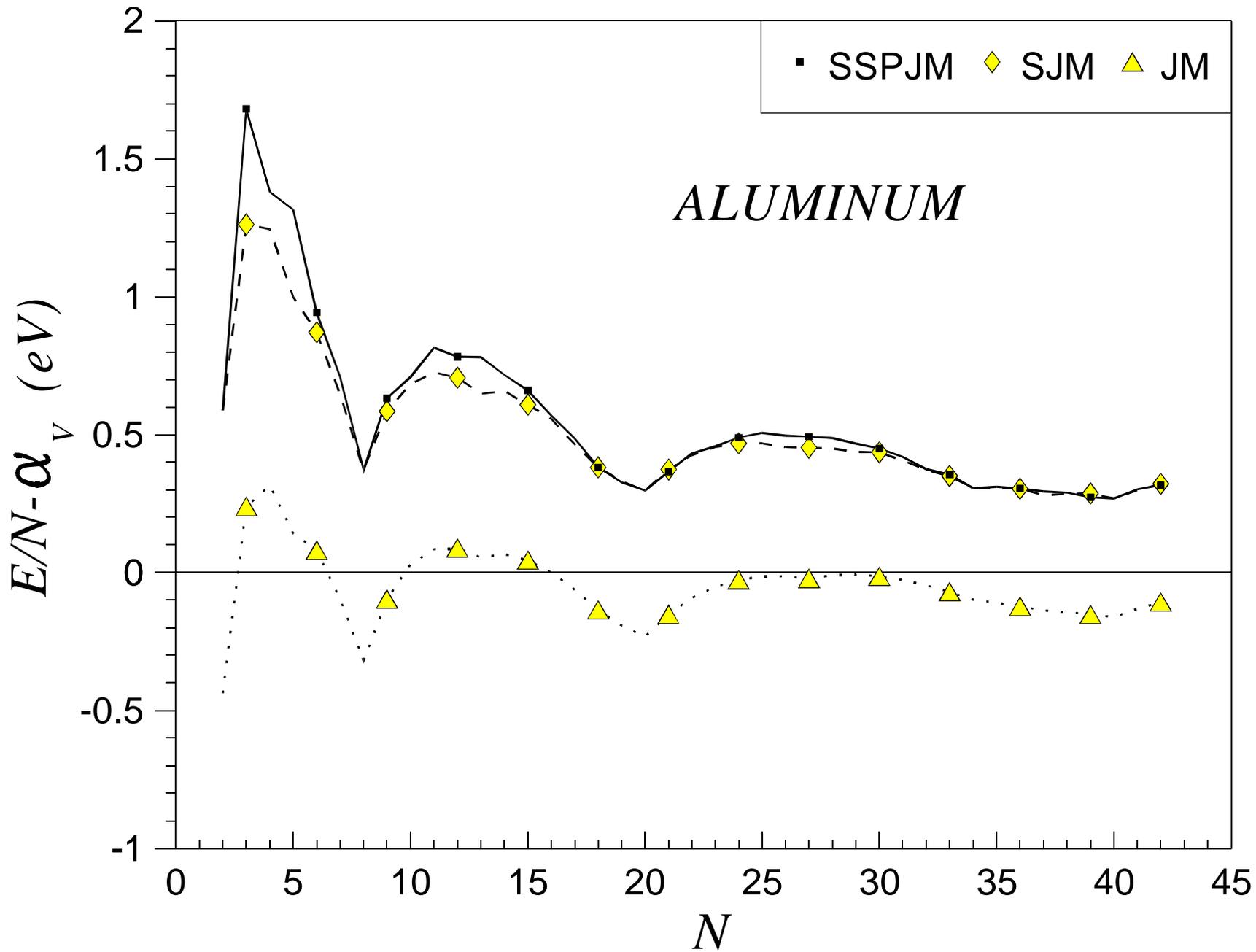
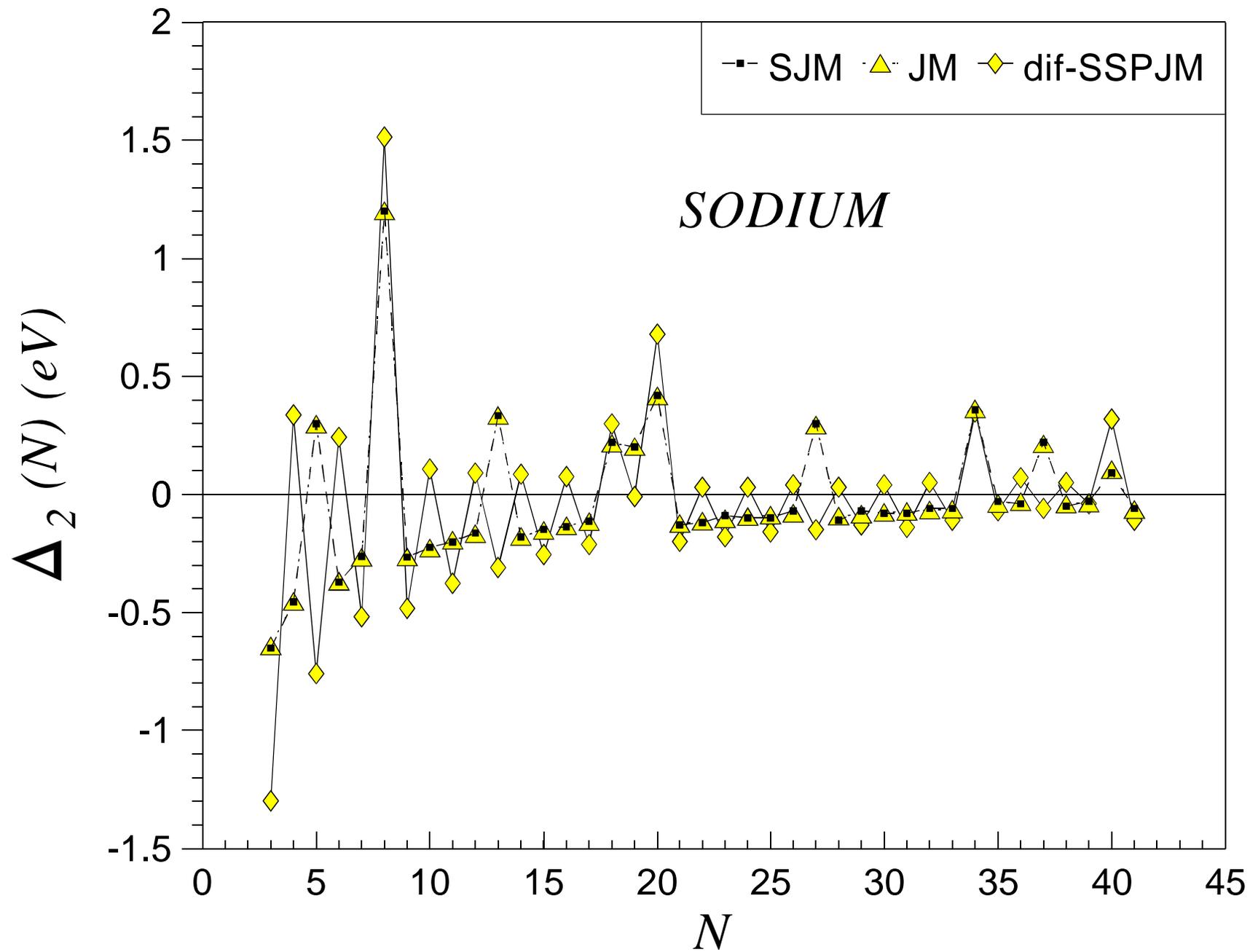
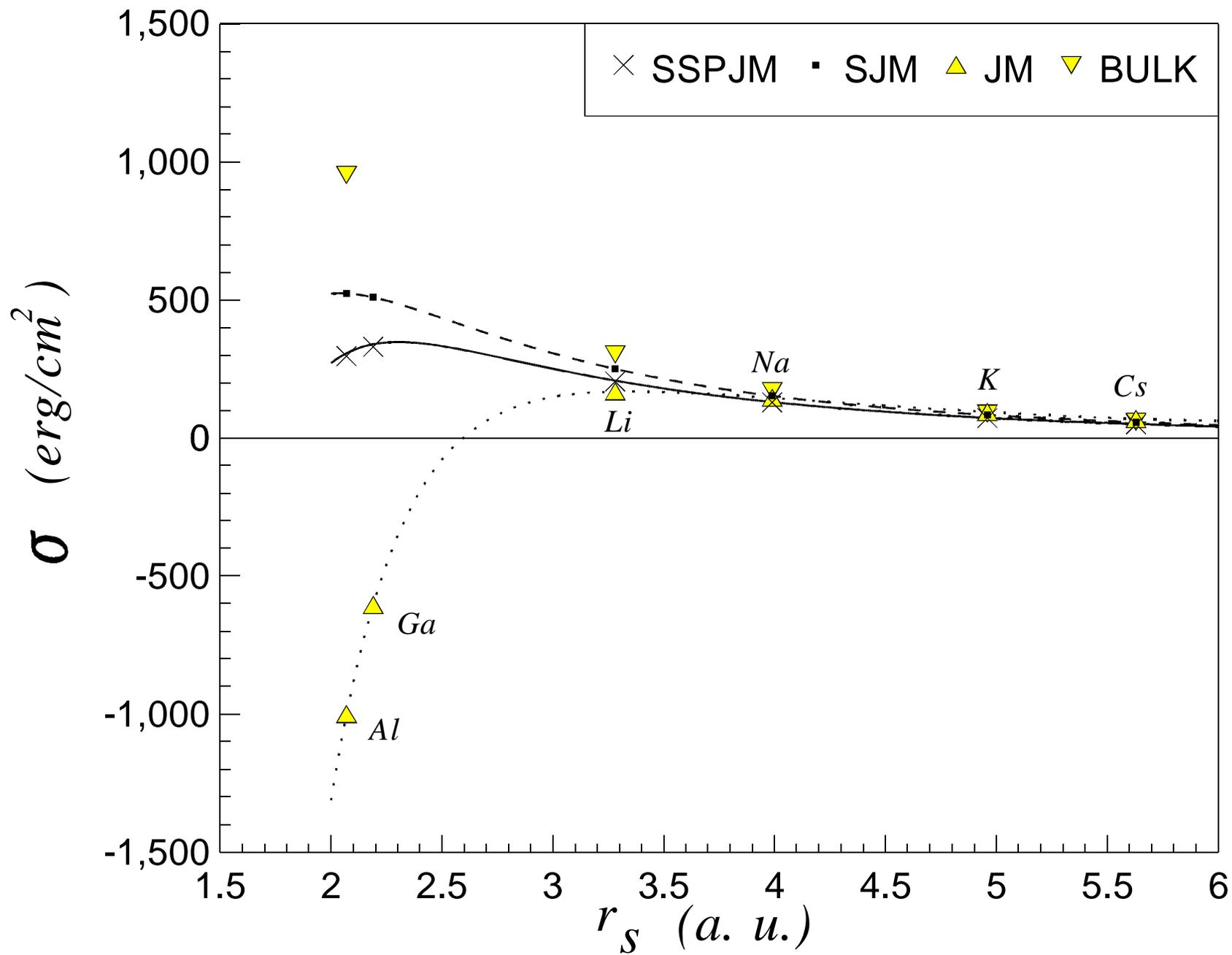
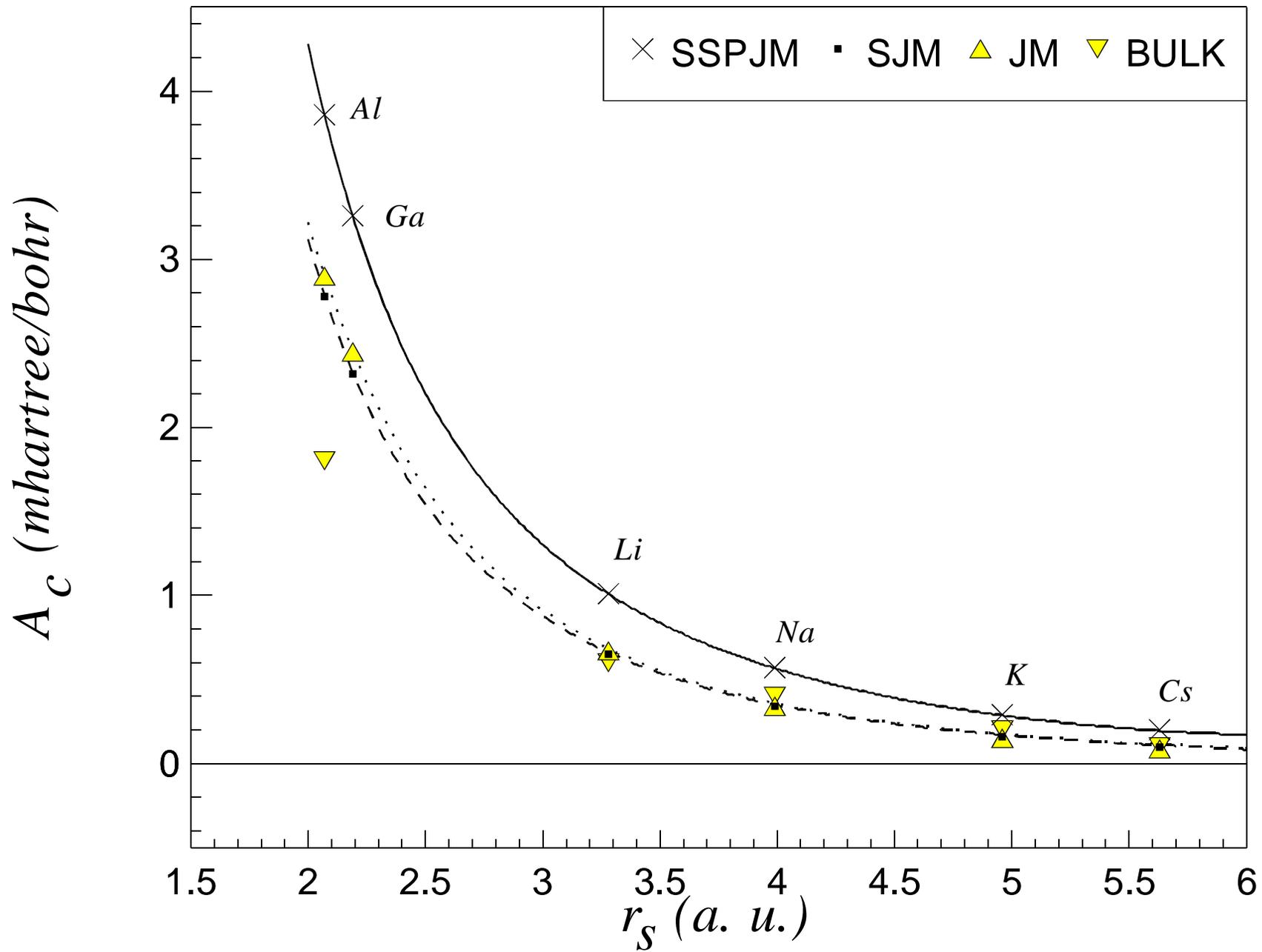
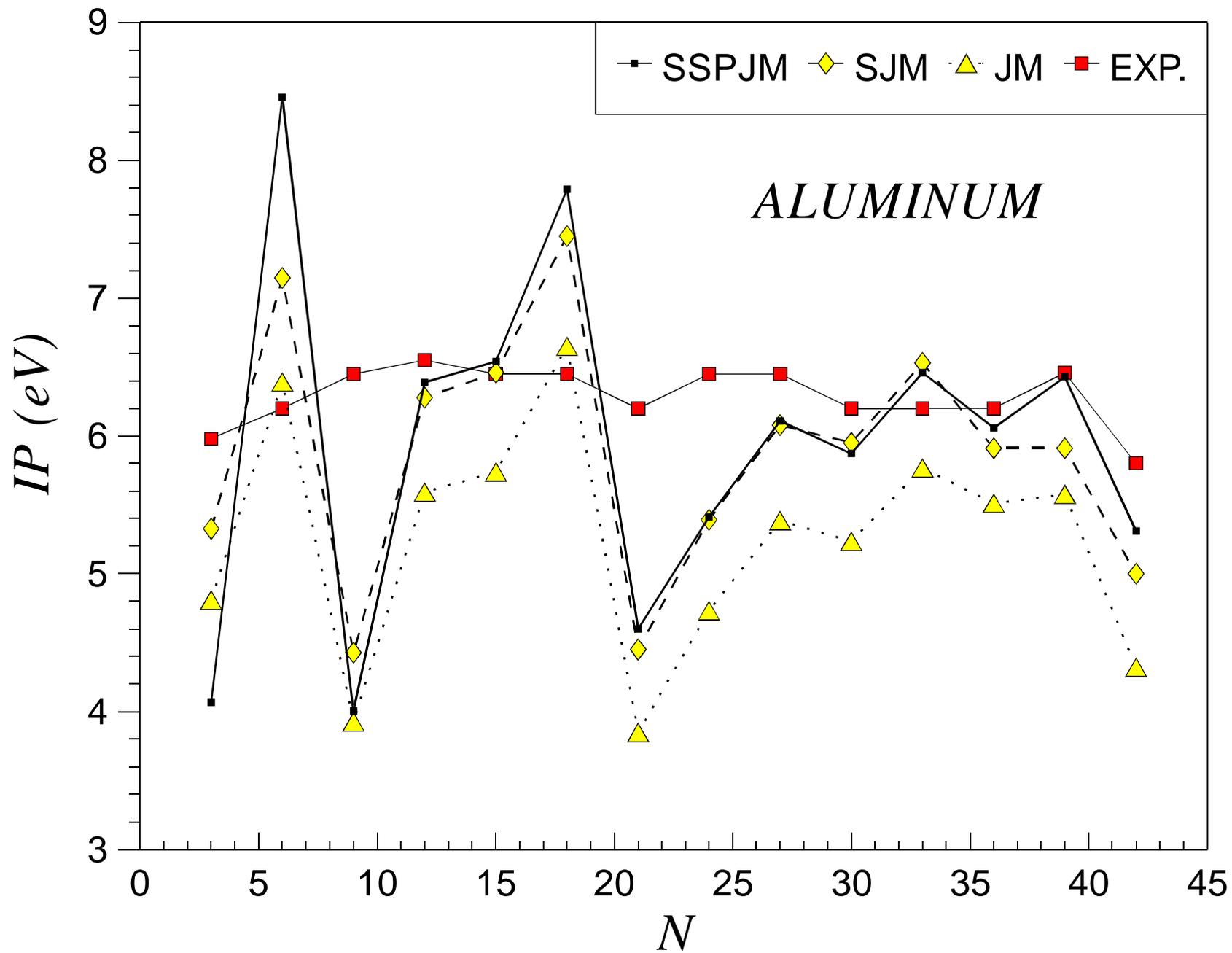
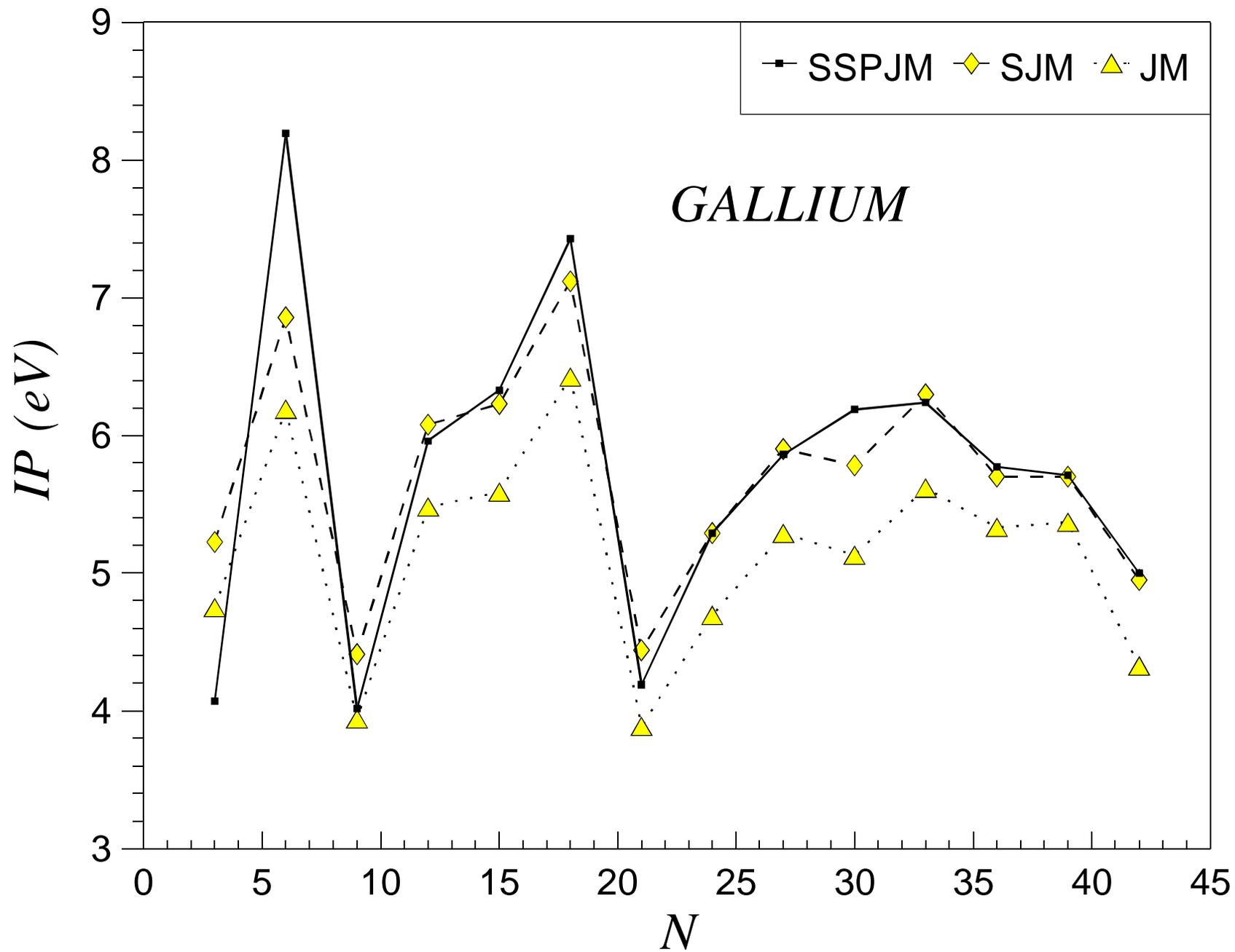
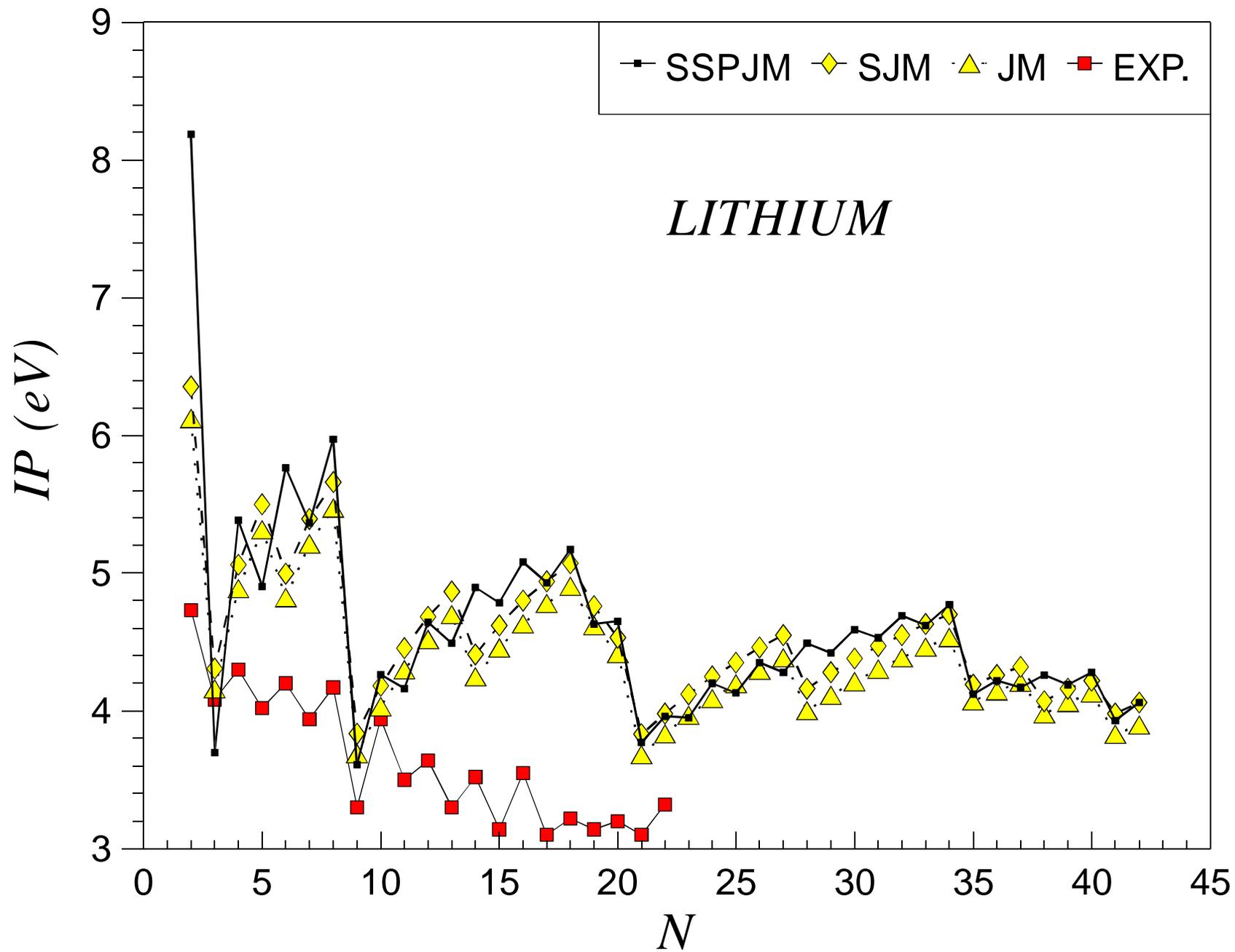
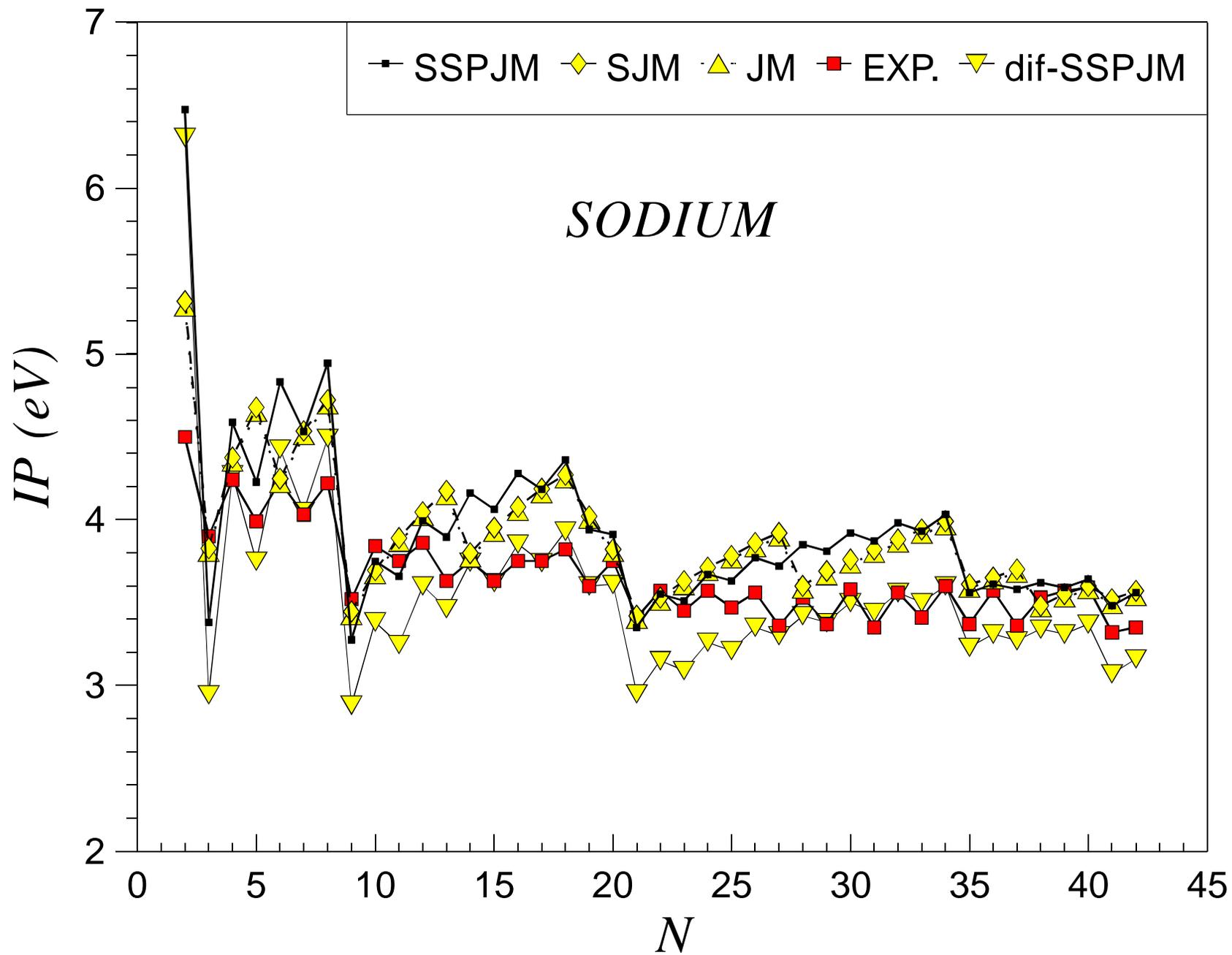
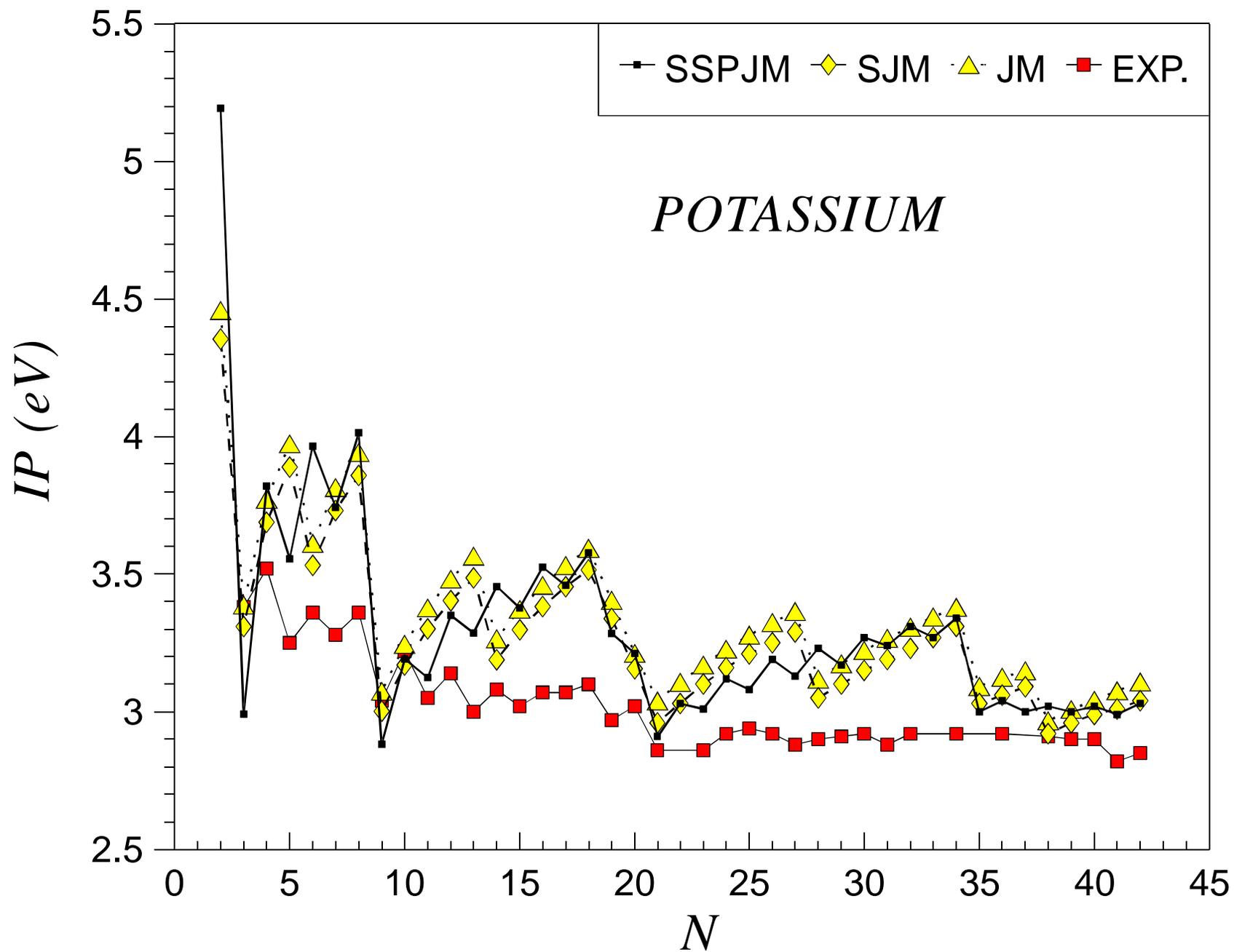
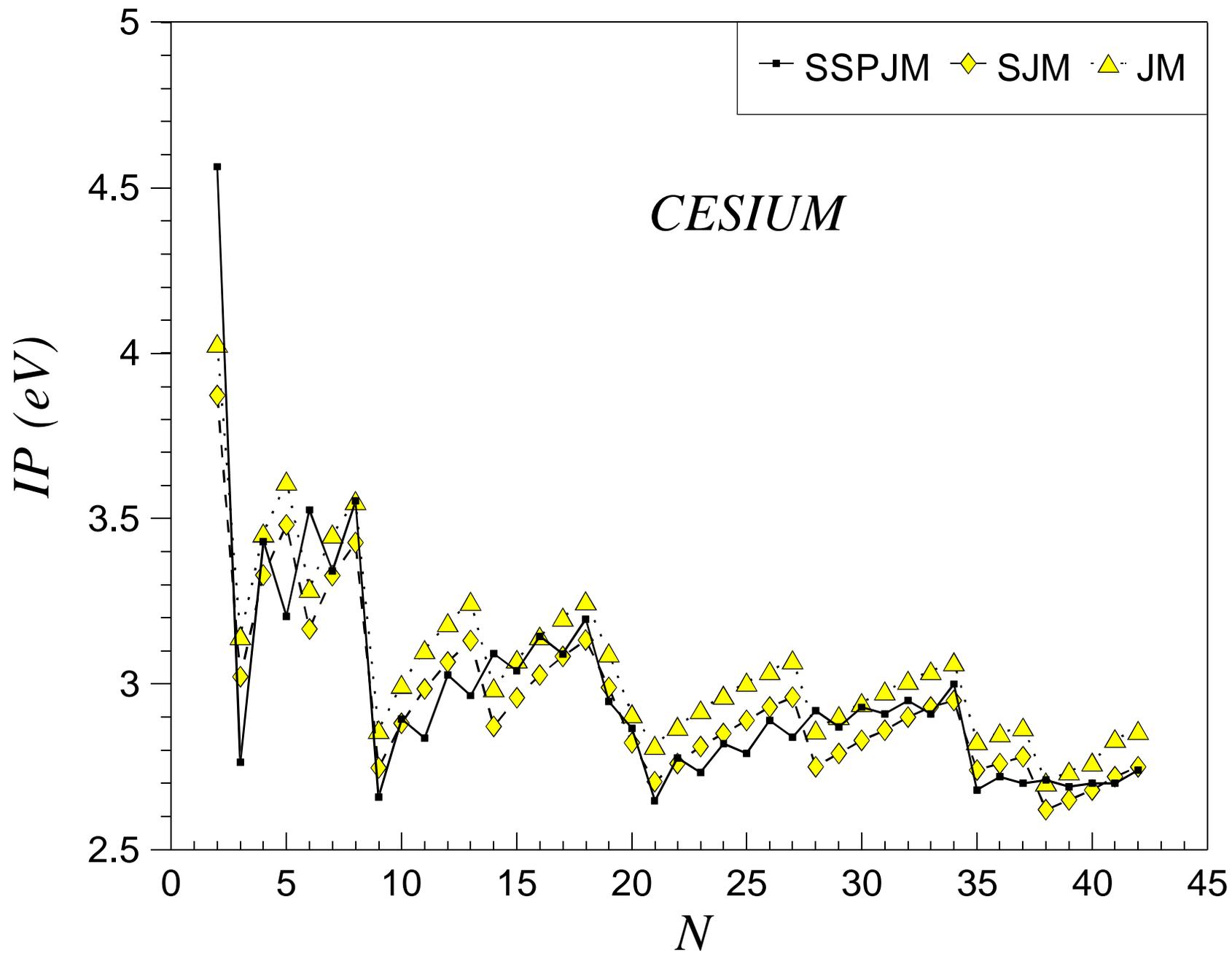
\end{document}